\documentclass[longauth]{aa}

\usepackage{txfonts}                    % tx fonts
\usepackage[utf8]{inputenc}     % allow utf-8 input (only pdflatex, lualatex etc.)
\usepackage{amsmath}                    % advanced maths commands
\usepackage{nccmath}                    % extends amsmath (ceqn environment etc.)

\usepackage[english]{babel}     % date formats, hyphenation, etc.
\usepackage{xcolor}                     % colors for hyperlinks
\usepackage{graphicx}                   % include figure files / images
\usepackage{booktabs}                   % midrule etc. (better than hline in tables)
\usepackage{gensymb}                    % Symbols like \degree

\usepackage[
colorlinks = true,
anchorcolor = blue,
linkcolor = blue,
urlcolor  = blue,
citecolor = blue,
]{hyperref}

\newcommand{\UY}{2003~UY$_\textrm{117}$}                    % not \UY117, digits NOT (!!) allowed
\newcommand{\Gaia}{\textit{Gaia~}}                          % Gaia should be written in slanted or italics style
\let\oldsim\sim
\renewcommand{\sim}{{\oldsim}}

\begin{document}

\title{Physical properties of trans-Neptunian object (143707) 2003~UY$_\textrm{117}$ derived from stellar occultation and photometric observations} % the 23 October 2020
\titlerunning{Physical properties of TNO (143707) 2003~UY$_\textrm{117}$}
%\subtitle{}

%
% Main authors
%
\author{
%
% Core (order: contributions and/or politics)
%
M.~Kretlow\inst{1}\and%\thanks{\scriptsize e-mail: mike@kretlow.de}\and
J.~L.~Ortiz\inst{1}\and
J.~Desmars\inst{2,3}\and
N.~Morales\inst{1}\and
F.~L.~Rommel\inst{10,5,7}\and
P.~Santos-Sanz\inst{1}\and
M.~Vara-Lubiano\inst{1}\and
E.~Fernández-Valenzuela\inst{7,1}\and
A.~Alvarez-Candal\inst{1,22}\and
R.~Duffard\inst{1}\and
F.~Braga-Ribas\inst{4,5}\and
B.~Sicardy\inst{3}\and
%
% Observer (O+), which are not already in core
%
A.~Castro-Tirado\inst{1}\and
E.~J.~Fernández-García\inst{1}\and
M.~Sánchez\inst{15}\and
A.~Sota\inst{1}\and
%
% Team or associated (order: alphabetical)
%
M.~Assafin\inst{9,5}\and
G.~Benedetti-Rossi\inst{14,5,6}\and
R.~Boufleur\inst{6}\and
J.~I.~B.~Camargo\inst{13,5}\and
S.~Cikota\inst{19}\and
A.~Gomes-Junior\inst{8,5}\and
J.~M.~Gómez-Limón\inst{1}\and
Y.~Kilic\inst{1,27}\and
J.~Lecacheux\inst{6}\and
R.~Leiva\inst{1}\and
J.~Marques-Oliveira\inst{6}\and
R.~Morales\inst{1}\and
B.~Morgado\inst{9}\and
%C.~Pereira\inst{13,5}\and
J.~L.~Rizos\inst{1}\and
F.~Roques\inst{6}\and
D.~Souami\inst{6,11,12}\and
R.~Vieira-Martins\inst{13,5,9}\and
%
% Observer (O-,no data,...; order: alphabetical)
%
%S.~Alonso\inst{23}\and
M.~R.~Alarcon\inst{21}\and
R.~Boninsegna\inst{26}\and
O.~Çakır\inst{30,31}\and
F.~Casarramona\inst{24}\and
J.~J.~Castellani\inst{25}\and
I.~de la Cueva\inst{16}\and
S.~Fişek\inst{28,29}\and
A.~Guijarro\inst{19}\and
T.~Haymes\inst{25}\and
E.~Jehin\inst{20}\and
S.~Kidd\inst{25}\and
J.~Licandro\inst{21}\and
J.~L.~Maestre\inst{17}\and
F.~Murgas\inst{21}\and
E.~Pallé\inst{21}\and
M.~Popescu\inst{18}\and
%F.~J.~Pozuelos\inst{20}\and
A.~Pratt\inst{25}\and
%A.~Roman\inst{15}\and
M.~Serra-Ricart\inst{21}\and
J.~C.~Talbot\inst{25}%\and
%
% Others
}

\authorrunning{M.~Kretlow et al.}

%
% List of institutions
%

\institute{
Instituto de Astrofísica de Andalucía, IAA-CSIC, Glorieta de la Astronomía s/n, 18008 Granada, Spain \and %1 (immer hier in main body)
Institut Polytechnique des Sciences Avancées IPSA, 94200 Ivrysur-Seine, France \and
IMCCE, Observatoire de Paris, PSL Research University, CNRS, Sorbonne Université, Univ. Lille, 75014 Paris, France \and
Federal University of Technology-Paraná (UTFPR/PPGFA), Curitiba, PR, Brazil \and
Laboratório Interinstitucional de e-Astronomia - LIneA - and INCT do e-Universo, Av. Pastor Martin Luther King Jr, 126 - Del Castilho, Nova América Offices, Torre 3000 / sala 817, Rio de Janeiro, RJ 20765-000, Brazil \and
LESIA, Observatoire de Paris, Université PSL, CNRS, Sorbonne Université, 5 place Jules Janssen, 92190 Meudon, France \and
Florida Space Institute, University of Central Florida, Orlando, FL 32826-0650, USA \and
Federal University of Uberlândia (UFU), Physics Institute, Av. João Naves de Ávila 2121, Uberlândia, MG 38408-100, Brazil \and
Universidade do Rio de Janeiro - Observatório do Valongo, Ladeira do Pedro Antonio 43, Rio de Janeiro, RJ 20.080-090, Brazil \and
Federal University of Technology - Paraná (PPGFA/UTFPR-Curitiba), Av. Sete de Setembro, 3165, Curitiba - PR - Brazil \and
Departments of Astronomy and of Earth and Planetary Science, 501 Campbell Hall, University of California, Berkeley, CA 94720, USA \and
naXys, Department of Mathematics, University of Namur, Rue de Bruxelles 61, 5000 Namur, Belgium \and
Observatório Nacional/MCTI, Rua Gal. José Cristino 77, Rio de Janeiro, RJ 20921-400, Brazil \and
UNESP-São Paulo State University, Grupo de Dinâmica Orbital e Planetologia, CEP 12516-410, Guaratinguetá, SP, Brazil \and
Sociedad Astronómica Granadina , Apartado de Correos 195, 18080 Granada, Spain \and
Agrupación Astronómica de Eivissa, C. Lucio Oculacio s/n, 07800 Ibiza, Spain \and
Observatorio Astronómico de Albox, Apt. 63, 04800 Albox, Almeria, Spain \and
Astronomical Institute of the Romanian Academy, 5 Cut¸itul de Argint, 040557 Bucharest, Romania \and
Centro Astronómico Hispano en Andalucía, Observatorio de Calar Alto, Sierra de los Filabres, E-04550 Gérgal, Almeria, Spain \and
University of Liège, Belgium \and
Instituto de Astrofísica de Canarias (IAC), C/Vía Láctea s/n, 38205 La Laguna, Tenerife, Spain \and
Instituto de F\'isica Aplicada a las Ciencias y las Tecnolog\'ias,U niversidad de Alicante, Alicante, San Vicente del Rapeig, Spain \and
Universidad de Granada, E.T.S. de Ingenierías Informática y de Telecomunicación, C/ Periodista Daniel Saucedo Aranda S/N, 18071 Granada, Spain \and
Agrupación Astronómica de Sabadell, Prat de la Riba sn, 08206 Sabadell, Spain \and
International Occultation Timing Association/European Section, Am Brombeerhag 13, 30459 Hannover, Germany \and
Rue de Mariembourg 45, 5670 Dourbes, Belgium \and
TÜBITAK National Observatory, Akdeniz University Campus, 07058 Antalya, Turkey \and
Department of Astronomy and Space Sciences, Faculty of Science, Istanbul University, 34116 Istanbul, Turkey \and
Istanbul University Observatory Research and Application Centre, 34116 Istanbul, Turkey \and
School of Mathematical and Physical Sciences, Macquarie University, Sydney, NSW 2109, Australia \and
Astrophysics and Space Technologies Research Centre, Macquarie University, Sydney, NSW 2109, Australia
}

\date{Received dd mmm, 2024; accepted dd mmm, 2024}

% \abstract{}{}{}{}{}
% 5 {} token are mandatory

\abstract
% context heading (optional). {} leave it empty if necessary
{Trans-Neptunian objects (TNOs) are considered to be among the most primitive objects in our Solar System. Knowledge of their primary physical properties is essential for understanding their origin and the evolution of the outer Solar System. In this context, stellar occultations are a powerful and sensitive technique for studying these distant and faint objects.}
%
% aims heading (mandatory)
{We aim to obtain the size, shape, absolute magnitude, and geometric albedo for TNO (143707) 2003~UY$_\textrm{117}$.}
%
% methods heading (mandatory)
{We predicted a stellar occultation by this TNO for 2020 October 23 UT and ran a specific campaign to investigate this event. We derived the projected profile shape and size from the occultation observations by means of an elliptical fit to the occultation chords. We also performed photometric observations of (143707) 2003~UY$_\textrm{117}$ to obtain the absolute magnitude and the rotational period from the observed rotational light curve. Finally, we combined these results to derive the three-dimensional shape, volume-equivalent diameter, and geometric albedo for this TNO.}
%
% results heading (mandatory)
{From the stellar occultation, we obtained a projected ellipse with axes of $(282 \pm 18) \times (184 \pm 32)$\,km. The area-equivalent diameter for this ellipse is $D_\mathrm{eq,A} = 228 \pm 21$~km. From our photometric $R$ band observations, we derived an absolute magnitude of $H_V = 5.97 \pm 0.07$\,mag using $V-R = 0.46 \pm 0.07$\,mag, which was derived from a $V$ band subset of these data. The rotational light curve has a peak-to-valley amplitude  of $\Delta m = 0.36 \pm 0.13$\,mag. We find the most likely rotation period to be $P = 12.376 \pm 0.0033$~hours. By combining the occultation with the rotational light curve results and assuming a triaxial ellipsoid, we derived axes of $a$ × $b$ × $c$ = $(332 \pm 24)$~km × $(216 \pm 24)$~km × $(180\substack{+28\\-24})$~km for this ellipsoid, and therefore a volume-equivalent diameter of $D_\textrm{eq,V} = 235 \pm 25$~km. Finally, the values for the absolute magnitude and for the area-equivalent diameter yield  a geometric albedo of $p_V = 0.139 \pm 0.027$.}
%
% conclusions heading (optional), leave it empty if necessary
{}


\maketitle

\section{Introduction}\label{sec:introduction}

Trans-Neptunian objects (TNOs), along with the objects coming from the Oort cloud, are considered to be the most primordial bodies in our Solar System. The study of their physical and dynamical properties helps us learn about their origin and evolution, which in turn provides crucial information about the origin and history of the early Solar System \citep{Nesvorny2012}. Currently, the Minor Planet Center (MPC) has counted about 5315 TNOs\footnote{\scriptsize Data retrieved from the MPC on 2024 January 22 .}. This includes the population of Centaurs, which are objects believed to be in a transition stage between TNOs and Jupiter-family comets \citep[e.g.,][]{Horner2004,Sarid2019}.

Because TNOs are located in the outer region of the Solar System, they are difficult to study. These objects typically exhibit low brightness, and with an average surface temperature of approximately $30-40$\,K, their thermal emission peak occurs in the far-infrared spectrum, a range obstructed by Earth's atmosphere. To derive radiometric sizes, it is necessary to observe them with space telescopes, as was done for more than 120 TNOs and Centaurs within the ESA \textit{Herschel} mission ``TNOs are Cool'' open-time key program \citep[e.g.,][and references therein]{Muller2009,Lellouch2013,Farkas-Takacs2020}. The \textit{Herschel} mission was completed in 2013.

An alternative to the radiometric technique for deriving sizes and albedos is the use of stellar occultations.
The observation of stellar occultations by small bodies (asteroids, comets, Centaurs, TNOs, and planetary moons) of the Solar System is an instrumental relatively simple, but powerful, technique for: directly measuring the size and shapes of these objects with (sub)kilometer accuracy, probing the environment around them with the possibility of revealing a binary nature \citep{Leiva2020}; discovering moons \citep[e.g.,][]{Gault2022} and rings \citep{Braga-Ribas2014,Ortiz2015,Ortiz2017}; and detecting, measuring, or constraining an atmosphere down to the nanobar pressure level \citep[e.g.,][]{Hubbard1988,Sicardy2003,Oliveira2022}. In addition, the occultation observation provides an astrometric measurement of the occulting object with (sub)milliarcsecond accuracy within the \Gaia reference system \citep[][]{Rommel2020,Ferreira2022,Kaminski2023}, which can be used to improve the orbit and therefore also predictions of future occultation events. In contrast to occultations by asteroids, predicting and successfully observing stellar occultations by TNOs can be very challenging, mainly due to the very small angular sizes of the TNOs together with the relatively large ephemeris uncertainties.

Trans-Neptunian object (143707) \UY{} was discovered by the 0.9 m Spacewatch telescope at Steward Observatory (Kitt Peak, Arizona) on 2003 October 23 (\href{https://minorplanetcenter.net/mpec/K03/K03V03.html}{MPEC 2003-V03}), shortly before its perihelion in the year 2004. Prediscovery observations date back to the year 2001 (\href{https://minorplanetcenter.net/mpec/K03/K03Y50.html}{MPEC 2003-Y50}). This TNO travels around the Sun in a highly eccentric orbit ($e ~\sim~ 0.4$), with a perihelion distance of $q ~\sim~ 32$\,au and an aphelion distance of $Q ~\sim~ 78$\,au. The MPC  classifies \UY{} as a scattered disc object (SDO), but not very much is known about its physical properties. \cite{Sheppard2012} obtained an absolute magnitude of $H_R = 5.35 \pm 0.03$\,mag and a color index of $V-R = 0.56 \pm 0.01$\,mag, and therefore $H_V = 5.91 \pm 0.04$\,mag. \cite{Alvarez-Candal2019} reported $H_V = 6.13 \pm 0.04$\,mag, $H_R = 5.60 \pm 0.04$\,mag, and $V-R = 0.53 \pm 0.06$\,mag.

The geometric albedo, $p_V$, and diameter, $D,$ of a small body are related via~\citep[e.g.,][]{Russell1916, Harris1998}
\begin{ceqn}\begin{align}\label{eq:albmag2diam}
D = \frac{D_0}{\sqrt{p_V}} 10 ^{-H_V/5},
\end{align}\end{ceqn}
where $H_V$ is the absolute magnitude of the object, $D_0 = 2\,\mathrm{au} \cdot 10^{V_\odot/5}$, and $V_\odot$ is the apparent visual magnitude of the Sun. Values for the apparent magnitude of the Sun are $V_{\odot} = -26.76$\,mag \citep{Willmer2018} and $V_{\odot} = -26.74$\,mag \citep{Rieke2008}, resulting in $D_0$ = 1330.2 km and $D_0$ = 1342.6 km, respectively. An earlier (commonly known and often used) value is $D_0$ = 1329 km. Applying Eq.~\ref{eq:albmag2diam}, with $D_0$ = 1330.2 km and assuming a geometric albedo, $p_V$, of either 6.9\% or 17\% for \UY{} as proposed by \cite{Santos-Sanz2012} for SDOs, yields effective diameters of about 333\,km and 212\,km, respectively. \cite{Farkas-Takacs2020} derived an effective diameter of $D = 196\substack{+114\\-54}$\,km for \UY{} from \textit{Herschel} (PACS\footnote{Photodetector Array Camera and Spectrometer}) thermal observations (using an absolute magnitude of $H_V=5.91$\,mag in their work).

In this paper, we report the observation of a stellar occultation by \UY{} and the results we obtained from it (Sect.~\ref{sec:occultation}). We also obtained photometric observations in order to derive  the absolute magnitude and %a rotational light curve from which we obtained
the rotation period of \UY{} (Sect.~\ref{sec:photometry}). Finally, we combined these results in order to constrain the three-dimensional size of the body and to derive the geometric albedo (Sect.~\ref{sec:results}).

\section{2020 October 23 occultation}\label{sec:occultation}

\subsection{Prediction}\label{sec:prediction}
\begin{figure}%[htb]
\includegraphics[width=\columnwidth]{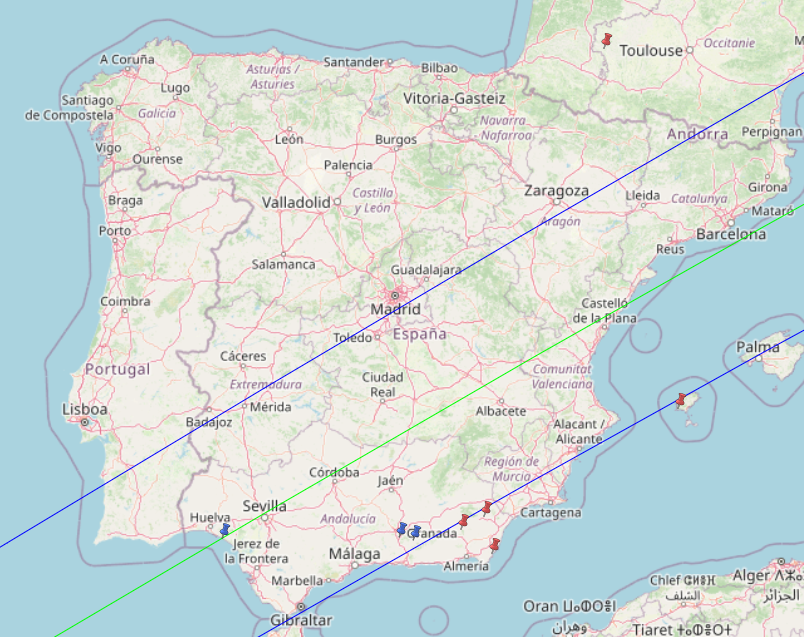}
\caption{Map of the ground track of our latest prediction update, based on astrometry obtained at the 2 m LT on La Palma. A spherical diameter of $D = 285$\,km was used for \UY{} for the predicted shadow path width (blue lines) plotted in this figure. The map also displays the sites where the occultation was observed, with blue markers indicating a positive detection and red ones indicating a negative detection (i.e., a "miss"). 
Negative observations reported from Belgium and England, which were located within the uncertainty of the original nominal prediction, are outside this map. Map credit: \href{https://www.openstreetmap.org}{OpenStreetMap}.
}\label{fig:prediction+observer}
\end{figure}

Within the Lucky Star collaboration\footnote{\scriptsize\url{https://lesia.obspm.fr/lucky-star/}}, we predicted a stellar occultation of a $G = 14.5$ mag star for 2020 October 23 using the \Gaia DR2 star catalog and the NIMA\footnote{Numerical Integration of the Motion of an Asteroid } ephemeris~\citep{Desmars2015}. Table~\ref{tab:occdata:2020} summarizes the occultation parameter and the details of the occulted star. The prediction details in Table~\ref{tab:occdata:2020} were taken from the nominal NIMA (version 3) prediction\footnote{\scriptsize \url{https://lesia.obspm.fr/lucky-star/occ.php?p=41240}}. The target star data from \Gaia DR3 are also given for comparison. About a week before the occultation date, we updated and refined the prediction using high-precision astrometry ($\sigma ~\sim~ 15$ mas) that we obtained with the 2\,m  Liverpool Telescope (LT) at Roque de Los Muchachos Observatory (ORM) on the island of La Palma, Spain. The update shifted the ground track farther to the north into a region with even better observability potential, especially for the European region, with a dense network of telescopes and observers (Fig.~\ref{fig:prediction+observer}). We then organized an observation campaign to detect the occultation from as many sites as possible. We used the Occultation Portal~\citep{Kilic2022}\footnote{\scriptsize\url{https://occultation.tug.tubitak.gov.tr/}} for observation reporting and data storage.

\begin{table}%[htb]
%\small
\centering
\caption{2020 October 23 occultation circumstances and target star data.}\label{tab:occdata:2020}
\begin{tabular}{ll}
\midrule\multicolumn{2}{c}{\textrm{Occultation parameter (NIMAv3 prediction)}}\\\midrule
Date and time of closest approach ($t_0$)                       &       2020-10-23\\
                                                                        &       22:18:08 UT $\pm 78$ s\\
Geocentric shadow velocity                              &       21.80 km/s\\
Magnitude drop                                          &       6.6 mag\\
Maximum duration                                                &   13.1 s\\
Apparent diameter of \UY                                &       12 mas \\[1em] %($D = 285$ km)
%Across-track error on sky plane                        &       1238 km \\

\midrule\multicolumn{2}{c}{\textrm{Occulted star data (from \Gaia DR2)}}\\\midrule
\Gaia DR2 source ID                                     &       62553763421986304\\
Proper motion (mas/yr)                          &       $\mu_\alpha* = +13.5 \pm 0.1$\\
                                        &       $\mu_\delta  = +0.5 \pm 0.0$\\
Position (ICRS, cat. epoch)                     &       $\alpha$ = 03 23 26.2108\\
                                                                        &       $\delta$ = +22 47 19.192\\
Position (ICRS, occ. epoch)                     &       $\alpha$ = 03 23 26.2160\\
                                                                        &       $\delta$ = +22 47 19.195\\
Position error (occ. epoch)                     &       $\sigma_{\alpha*} = 0.3 \:\textrm{mas}$\\
                                                                        &       $\sigma_\delta = 0.3 \:\textrm{mas}$\\
G, RP, BP magnitudes                                    &       14.54, 13.80, 15.17\\
V, R, B magnitudes (from NOMAD)         &       14.54, 14.27, 15.25\\
J, H, K magnitudes (from NOMAD)                 &       12.84, 12.41, 12.22\\
\midrule\multicolumn{2}{c}{\textrm{Occulted star data (from \Gaia DR3)}}\\\midrule
%Gaia DR2 source ID                                     &       62553763421986304\\
Proper motion (mas/yr)                          &       $\mu_\alpha* = +13.320 \pm 0.049$\\
                                        &       $\mu_\delta  = +0.388 \pm 0.029$\\
Position (ICRS, cat. epoch)                     &       $\alpha$ = 03 23 26.2113\\
                                                                        &       $\delta$ = +22 47 19.192\\
Position (ICRS, occ. epoch)                     &       $\alpha$ = 03 23 26.2155\\
                                                                        &       $\delta$ = +22 47 19.194\\
Position error (occ. epoch)                     &       $\sigma_{\alpha*} = 0.24 \:\textrm{mas}$\\
                                                                        &       $\sigma_\delta = 0.14 \:\textrm{mas}$\\
%G, RP, BP magnitudes (from DR3)                &       14.53, 13.79, 15.14\\

RUWE                                                                    &         1.05 \\
Duplicated source                                       &       false\\
\midrule
\end{tabular}
\tablefoot{The maximum occultation duration (central line) and the apparent diameter of the TNO are given for an assumed size of $D = $~285 km. The $V,R,B,J,H$, and $K$ magnitudes of the target star were taken from the NOMAD catalog~\citep{Zacharias2004}. The renormalized unit weight error (RUWE) is a measure of the reliability of a single-star model derived from observations. A value close to 1 is typically expected. Values exceeding 1.4 could indicate that the source is not a single star or that there
are problems with the astrometric solution. The duplicated source flag also indicates that multiple sources were obtained during data processing \footnote{\scriptsize\url{https://gea.esac.esa.int/archive/documentation/GEDR3/Gaia_archive/chap_datamodel/sec_dm_main_tables/ssec_dm_gaia_source.html}}.}
\end{table}

\subsection{Observations}\label{sec:observations}
\begin{figure}%[tb]
\centering
\includegraphics[width=\columnwidth]{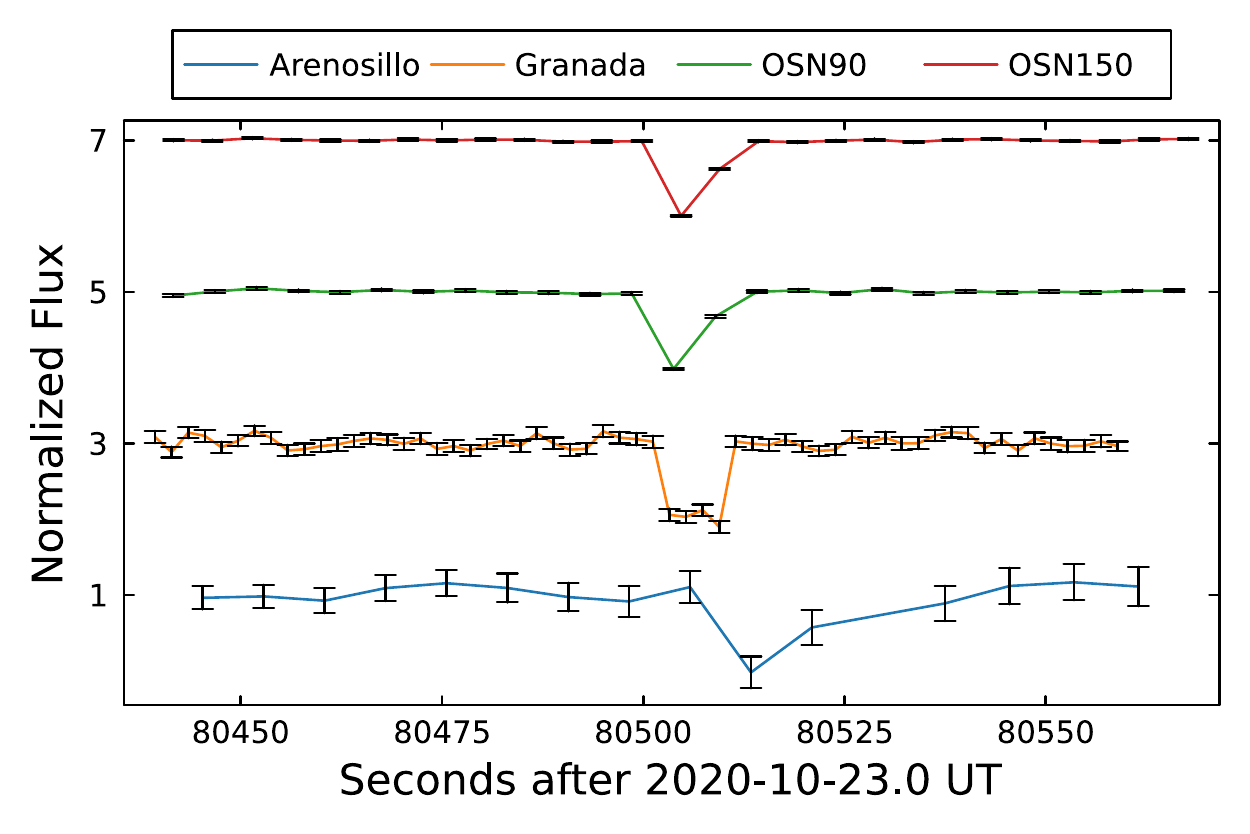}
\caption{Occultation light curves. The light curves (flux vs. time) are normalized and shifted from each other on the y-axis by an offset value of 2 for clarity. The site, telescope, and instrument details are given in Table~\ref{tab:obs2020}.}\label{fig:olc_all}
\end{figure}

The weather conditions were unfavorable during the event for large parts of the occultation path. However, we obtained four positive detections from three different locations in Spain. Additionally, we recorded three very close misses to the south of the body from Spain and ten misses from another nine observing sites (see Fig.~\ref{fig:prediction+observer} and Table~\ref{tab:obs2020} for observation details). We utilized synthetic aperture photometry to obtain the occultation light curves from the observations. The four positive detections are shown in Fig.~\ref{fig:olc_all}.

The star's apparent diameter was calculated to be 0.0178 mas ($V$-mag) and 0.0172 mas ($B$-mag) using the formulae published by \cite{Kervella2004}. This translates to a distance of 0.4\,km at the projected distance of \UY{} ($\Delta = 33.47$ au), or 0.02\,s for the shadow velocity of 21.80\,km/s. The Fresnel scale $R_F = \sqrt{\lambda \cdot \Delta / 2}$ is 1.32\,km, or 0.06\,s for a wavelength of $\lambda = 700$\,nm. As all positive detections were recorded with exposure times $\ge 2$\,s, any effects due to diffraction or the apparent stellar diameter are negligible.

The ingress (disappearance) and egress (reappearance) times were extracted from the fitted occultation light curves and were translated into chords on the sky plane. To model the light curves and to fit the profile, we utilized the SORA\footnote{Stellar Occultation Reduction and Analysis} Python package \citep{Gomes-Junior2022}, which also facilitates the extraction of ingress and egress times. The extracted times are listed in Table~\ref{tab:tim2020}.

\begin{table*}%[tb]
\caption{Ingress and egress times (UT) obtained for the 2020 October 23 occultation.}\label{tab:tim2020}
\centering
\begin{tabular}{@{\extracolsep{\fill}}rlllllrc@{}}
\toprule\midrule
\# & Site name     & Ingress (UT)       & I.err         & Egress (UT)   & E.err           & Duration   & Chord length \\\midrule
1  & Arenosillo    & 22:21:49.83        & 0.69\,s       & 22:22:00.44   & 0.54\,s         & 10.61\,s      & 231\,km\\ %231.3
2  & Granada       & 22:21:42.36        & 0.14\,s       & 22:21:50.48   & 0.08\,s         &  8.12\,s      & 177\,km\\ %177.0
3  & OSN90         & 22:21:41.12        & 1.25\,s       & 22:21:48.45   & 0.23\,s         &  7.34\,s      & 160\,km\\ %160.0
4  & OSN150        & 22:21:42.18        & 1.03\,s       & 22:21:49.13   & 0.26\,s         &  6.94\,s      & 151\,km\\ %151.2
\midrule
\end{tabular}
\tablefoot{Also given are the $1\sigma$ errors of the ingress and egress times, the occultation duration in seconds, and the corresponding chord length in km.}
\end{table*}

\subsection{Profile fit}
\begin{figure}%[htb]
\centering
\includegraphics[width=\columnwidth]{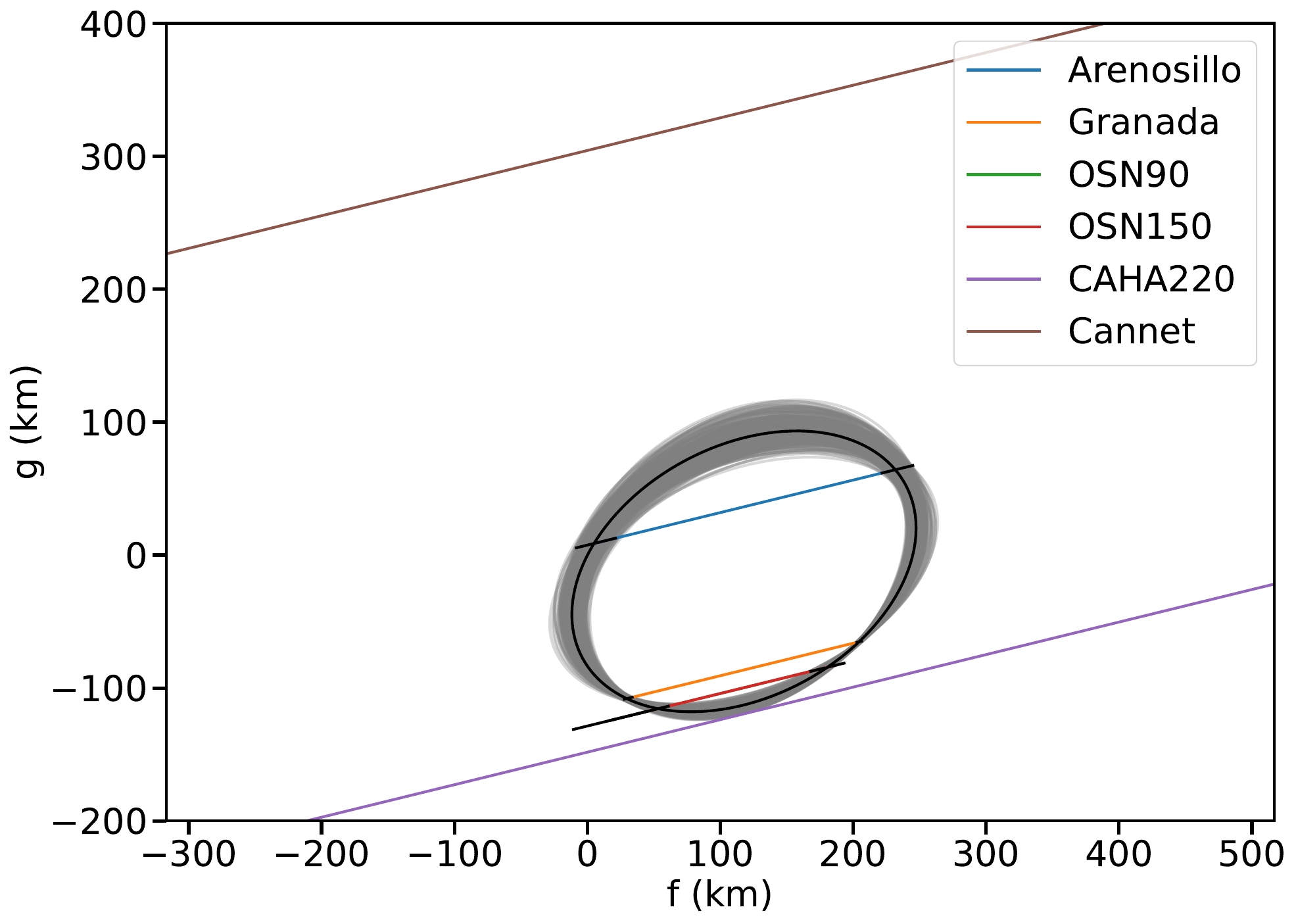}
\caption{Elliptical fit to the 2020 October 23 occultation observations (chords). This fit describes the limb of \UY{} for the moment of the occultation on the sky plane, defined by the $(f,g)$ axes. As two chords were derived from the same site (OSN90 and OSN150; Sierra Nevada Observatory), they are not distinguishable in the plot. Also shown is the chord for the nearest site to the south (CAHA220: Calar Alto Observatory, 2.2\,m telescope) and to the north (Cannet), both of which had a negative detection ("miss"). 
%A second site at the southern border with a miss, which has an almost identical positioned chord as CAHA220, is Albox. 
The gray shaded area is the $1\sigma$ uncertainty region of the derived ellipse.}
\label{fig:20201023_profile_fit}
\end{figure}

Assuming a spheroidal or an ellipsoidal object, the projected cross section on the sky plane is an ellipse. Therefore, we fit an ellipse to the extremities of the chords (derived from the ingress and egress times as described in Sect.~\ref{sec:observations}), taking into account the near misses of the occultation as an additional constraint (Fig.~\ref{fig:20201023_profile_fit}). The five solve-for parameters were: the center of the ellipse $(f,g)$ with respect to the center of the fundamental plane defined by the geocentric star and TNO position for the event time; the semimajor axis, $a'$; the oblateness, $\epsilon' = (a'-b')/a'$; and the position angle of the ellipse, $\varphi'$\footnote{\scriptsize The (clockwise positive) angle between the ``g-positive'' direction (i.e., north) and the semiminor axis, $b'$.}. The prime ($'$) indicates that these parameters belong to the projected ("apparent") profile ellipse of the object and distinguishes them from the axes of a physical body (triaxial ellipsoid with semiaxes $a,b$, and $c$). The parameters were estimated using the Levenberg-Marquardt optimization algorithm. The goodness of the fit was evaluated from the $\chi^2$ per degree of freedom (pdf) value, defined as $\chi^2_\textrm{pdf} = \chi^2 / (N-M)$, where $N=8$ is the number of data points and $M=5$ is the number of adjustable parameters. Ideally, this value should be close to one. We obtain  $\chi^2_\textrm{pdf} = 0.43$ for our fit. The $1\sigma$ uncertainties in the retrieved parameters were obtained from a grid search in the parameter space, by varying one parameter from its nominal solution value while keeping the other parameters constant. Acceptable values were those that gave a $\chi^2$ between $\chi^2_\textrm{min}$ and $\chi^2_\textrm{min} + 1$. The results of our instantaneous best-fitting limb are summarized in Table~\ref{tab:profile_fit_result}.

\begin{table}%[!htb]
\caption{Elliptical occultation limb profile fit result.}\label{tab:profile_fit_result}
\centering
\begin{tabular}{lr}%@{\extracolsep{\fill}}lr@{}}
\toprule\midrule
Center coordinates (f,g)                                        &       $(119 \pm 7, -10 \pm 8)$ km \\
Semi-major axis $a'$                                            &       $141 \pm 9$ km \\
Semi-minor axis $b'$                                            &       $94 \pm 16$ km\\
Position angle $\varphi'$                                       &       $-31 \pm 11$ deg\\
Oblateness $\epsilon'$                                          &       $0.33 \pm 0.11$\\
Area-equiv. diameter $D_\textrm{eq,A}$          &       $228 \pm 21$ km\\
Best-fit $\chi^2_\textrm{pdf}$                          &   0.43\\
\midrule
\end{tabular}
\tablefoot{The fitted ellipse center (f,g) is with respect to the JPL\#20 Horizons ephemeris.
%$\chi^2_\textrm{pdf} = \chi^2 / (N-M)$ where $N=8$ are the number of data points and $M=5$ is the number of adjustable parameters.
}
\end{table}

\section{Photometry}\label{sec:photometry}
To interpret the occultation results with respect to the three-dimensional shape and size of the physical body, we carried out photometric observations of \UY{} to determine its rotational light curve.
We carried out observations with the 1.5\,m telescope at Sierra Nevada Observatory (Spain) and with the 1.23\,m telescope at Calar Alto Observatory (Spain) over six nights and with longer time coverage than the observations with the 2\,m LT on La Palma (Sect.~\ref{sec:prediction}); the latter were done with the IO:O instrument and a Sloan $r'$ filter, and  were focused on astrometry with the aim of updating the occultation prediction. We also used sparse observations made at Calar Alto in 2019 by our group. Observations at the 1.5\,m telescope were made with an Andor iKon-L CCD camera (model DZ936N-BEX2-DD)\footnote{\scriptsize \url{https://www.osn.iaa.csic.es/en/page/ccdt150-and-ccdt90-cameras}} without filters in order to increase the signal-to-noise ratio. At the 1.23\,m telescope, we used the DLR-MKIII instrument\footnote{\scriptsize  \url{https://www.caha.es/es/telescope-1-23m-2/ccd-camera}} (also without filters), except for two nights when we took images with $V$ and $R$ filters to determine the $V-R$ color of the TNO. Image calibration and photometry were performed using the same algorithms and procedures as for the LT images (Sect.~\ref{sec:prediction}). The science images were calibrated in the usual manner, namely, bias and flat field image correction were applied to them.

\subsection{Absolute magnitude}\label{sec:absmag}
\begin{figure}%[htb]
\includegraphics[width=\columnwidth]{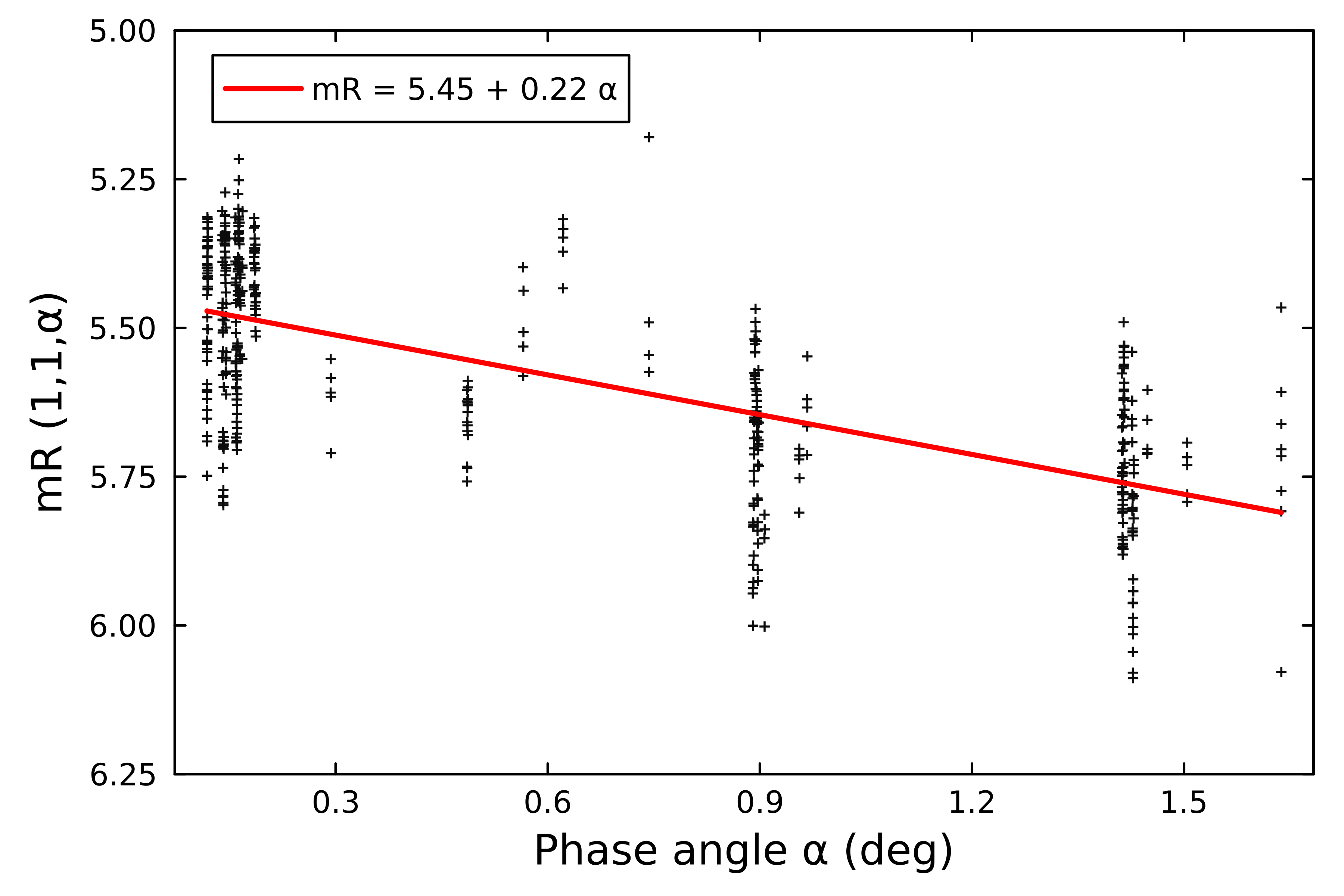}
\caption{Reduced magnitude $m_R(1,1,\alpha)$ vs. the phase angle, $\alpha$. In total, 431 observations obtained with the 2\,m LT, with the 1.5 m telescope at Sierra Nevada Observatory, and with the 1.23\,m telescope at Calar Alto Observatory were used. The brightness variation cluster is due to the rotation of \UY{}.}\label{fig:R11vsPHA}
\end{figure}

From the calibrated CCD images, we derived magnitudes in the $R$ band using our algorithms that use the \Gaia DR2 field stars to determine photometric transformation equations, which take the color information into account~\citep{Morales2022}. For the color of the TNO, we used a value of $V-R = 0.46 \pm 0.07$\,mag, which was derived from two nights of observations with the Calar Alto 1.23\,m telescope.  From a linear regression of the reduced $R$~magnitude (the apparent magnitude that the TNO would have at 1\,au from the Sun and Earth) at several phase angles, $\alpha$, we obtained the absolute magnitude, $H_R \equiv m_R(1,1,0),$ and the phase slope, $\beta$ (Fig.~\ref{fig:R11vsPHA}). From the fitted trend line, we derived an absolute magnitude of $H_R = 5.45 \pm 0.01$\,mag  and a slope parameter of $\beta = 0.22 \pm 0.01$\,mag/$\degree$. Our observations covered the phase angle range $\alpha = [0.118\degree,1.638\degree]$. The scatter around the trend line indicates a significant rotational modulation.

\subsection{Rotational light curve}
\begin{figure}%[htb]
\includegraphics[width=\columnwidth]{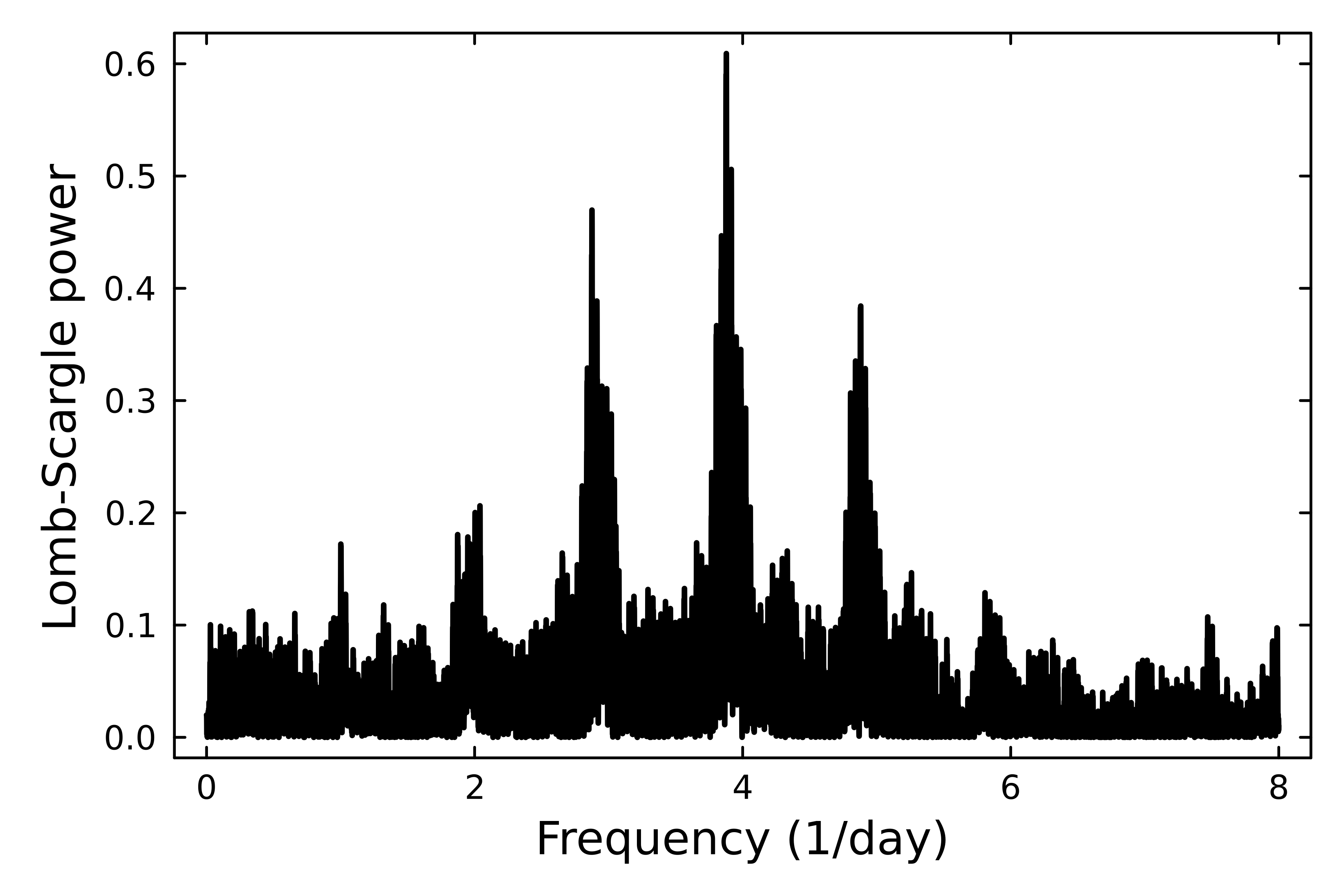}
\caption{LS periodogram. The rotational light curve frequency is given in 1/day. These values are light curve frequencies (periods); since a small body typically executes two light curve periods in a single rotation period, the best rotation period we obtain from the LS analysis is $P=12.3765$\,h.}\label{fig:LSperiodogram}
\end{figure}

After de-trending\footnote{i.e., using the O-C residuals of the linear fit described in Sect.~\ref{sec:absmag}.} the reduced and light-travel-time-corrected photometry, we performed a search for the rotation period of \UY{} using different period-finding techniques. In total, we had more than 400 observations from 2019 November 30 to 2021 February 15. We used the Lomb-Scargle (LS; \cite{Lomb1976}; \cite{Scargle1982}) algorithm to estimate the most likely rotation period from our data. The algorithm works in the frequency domain, and the most prominent light curve frequencies (in units of 1/day as the timescale of the data is days) are shown in the LS periodogram (Fig.~\ref{fig:LSperiodogram}). The normalized spectral power reveals a frequency of about 4/day (the exact value is $f = 3.878329$) as the dominant frequency. Given that a small (ellipsoidal) body typically executes two light curve periods in a single rotation period, the most likely rotation frequency is $f = 3.878329 / 2$ (day$^{-1}$), which corresponds to a rotation period of $P = 12.3765$\,h. 

\begin{figure}%[htb]
\includegraphics[width=\columnwidth]{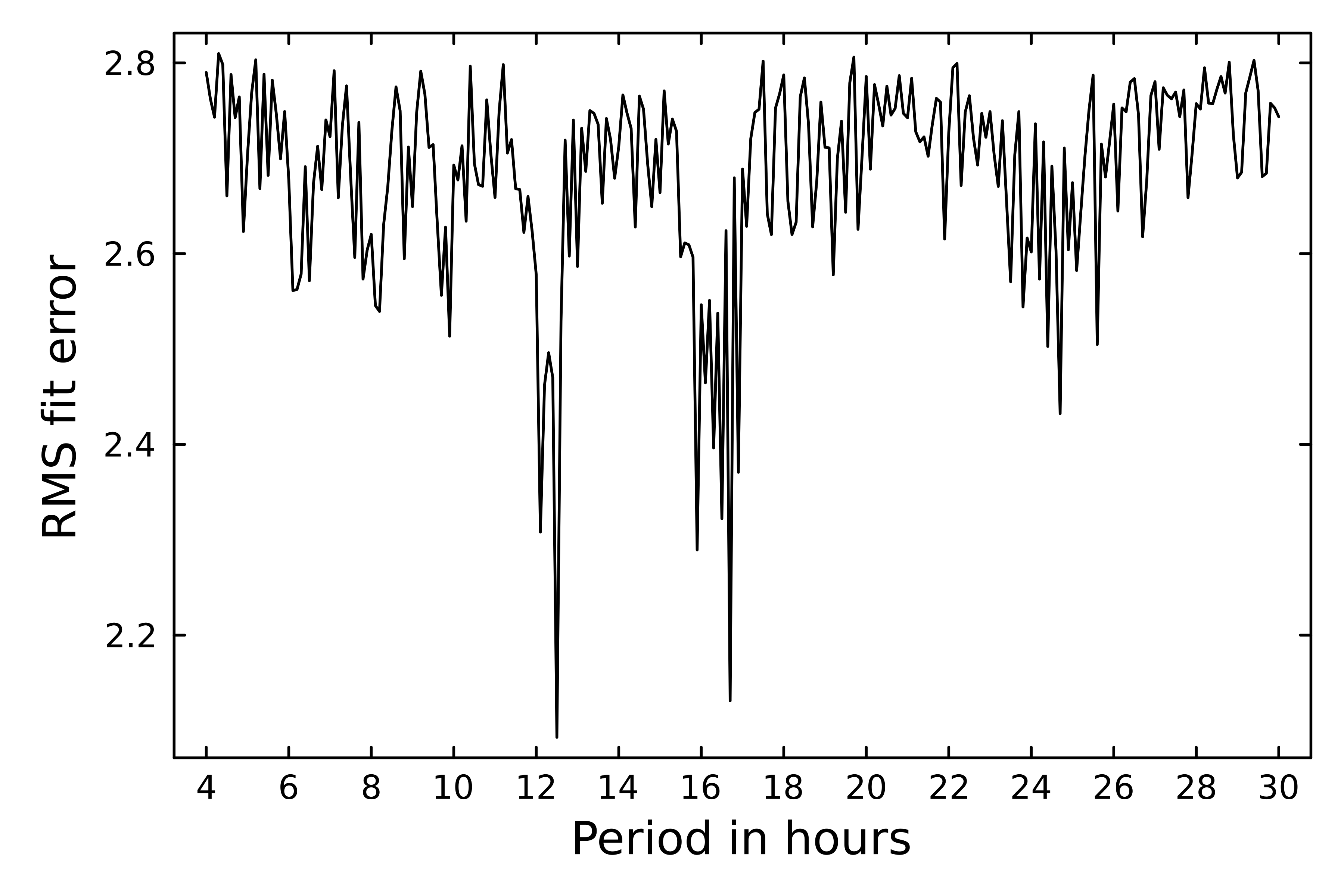}
\caption{Fourier fit periodogram. We scanned the period range  $4-30$ hours in 0.00001\,h steps, folded the photometric data with the selected period, and fit a second-order Fourier series to the data. The best period we obtained, $P=12.3763$\,h, is the value for which the RMS of the residuals (data minus fit) reaches a minimum.}\label{fig:FouFitPeriodogram}
\end{figure}

\begin{figure}%[htb]
\includegraphics[width=\columnwidth]{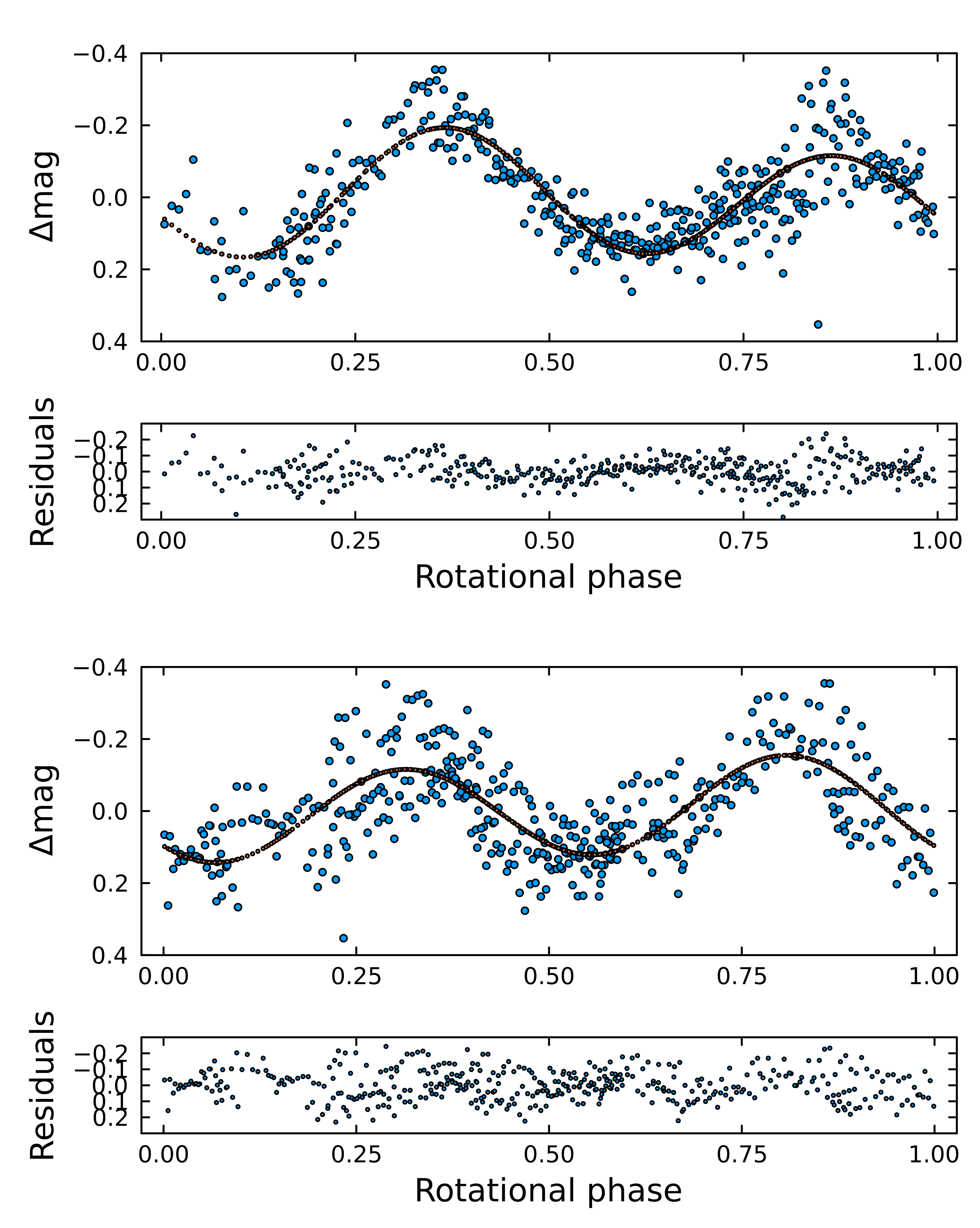}
\caption{Phased rotational light curve for \UY{} using all the photometric data folded with a period of $P=12.376$\,h (upper panel) and $P=16.692$\,h (lower panel). 
The $P=12.376$\,h is our preferred solution. This double-peaked light curve with an amplitude of 0.36\,mag indicates a highly nonspherical body with a presumable triaxial shape. Phase 0.0 corresponds to the moment of mid-occultation for the observed occultation chords.}\label{fig:PhasedRLC}
\end{figure}

To further constrain and verify the rotation period, we folded the data with a period in the range 4\,h to 30\,h in 0.00001 day steps. For each period value, we fit a second-order Fourier series to the folded data and calculated the root mean square error  (RMSE; Fig.~\ref{fig:FouFitPeriodogram}). Additionally, we verified the phased plots for the most prominent periods that were derived in both approaches, and we also evaluated split-halves plots. The period with the smallest RMSE is $P = 12.3763$\,h, which we chose as our best estimate for the rotation period of \UY{}. The second prominent minimum at 16.692\,h (Fig.~\ref{fig:FouFitPeriodogram}) corresponds to 2.8756 cycles/day, which is a 24\,h alias of the 1.9392 cycles/day frequency that corresponds to our preferred period of 12.376\,h. The periodogram in cycles/day (Fig.~\ref{fig:LSperiodogram}) in addition to the periodogram in hours helps to identify 24\,h aliases of the main peak, which are usually separated by $\sim$~1 cycle/day. From the results of the two approaches, we conclude that the rotation period is $P = 12.376 \pm 0.0033$\,h\footnote{The 1$\sigma$ error was derived from the period value differences corresponding to $\chi^2_\textrm{min}$ and $\chi^2_\textrm{min} + 1$.}. From the best Fourier fit (Fig.~\ref{fig:PhasedRLC}), we derived a peak-to-valley amplitude of $\Delta m = 0.36 \pm 0.13$\,mag.

\section{Results}\label{sec:results}
\subsection{Size and shape}
The instantaneous occultation ellipse limb fit to the projected profile of \UY{} (Fig.~\ref{fig:20201023_profile_fit}) yields $a' = 141 \pm 9$ km, $b' = 94 \pm 16$ km, and $\varphi' = -31\degree \pm 11\degree$ with a $\chi^2_\textrm{pdf} = 0.43$ value   (Table~\ref{tab:profile_fit_result}).
Given the large amplitude of the rotational light curve, we can expect that the intensity variations are due to shape effects, and this implies that \UY{} is not a spheroid. We assumed that a triaxial ellipsoid with semiaxes $a>b>c$ (spin-axis $c$) is a good approximation for the physical body. In the following, we deduce the possible shape and size of this ellipsoid from the occultation observation combined with our light curve results. Using the Maclaurin sequence equations \citep{Chandrasekhar1969}, a hydrostatic equilibrium body rotating at $\sim~ 12$\,h down to a density of $\rho ~\sim~ 0.2$ g/cm$^3$ would have taken on a Maclaurin spheroid shape, but \UY{} clearly does not have an (oblate) spheroid shape. Valid triaxial ellipsoids in hydrostatic equilibrium (Jacobi solutions) are possible only for densities ranging from 0.254 g/cm$^3$ to 0.333 g/cm$^3$ given the rotation period of 12.38\,h. These densities are too low to be realistic for bodies in this size range.

The orthogonal projection of a triaxial ellipsoid (axes $a>b>c$, rotating around $c$) for a given spin state, expressed by the aspect angle, $\psi$ (i.e., the angle between the rotation axis, $c,$ and the line of sight) and the rotational phase, $\phi$, is \citep[e.g.,][]{Magnusson1986}
\begin{align}
A  &= b^2c^2 \sin^2\psi \sin^2\phi + a^2c^2 \sin^2\psi\cos^2\phi + a^2b^2\cos^2\psi,\\
-B &= a^2(\cos^2\psi\sin^2\phi + \cos^2\phi) + b^2(\cos^2\psi\cos^2\phi + \sin^2\phi) \notag \\&\quad + c^2\sin^2\psi,\\
a' &= \left ( \frac{2A}{-B-(B^2-4A)^{1/2}} \right )^{1/2} \label{eq1:magnusson1986} ,\\
b' &= \left ( \frac{2A}{-B+(B^2-4A)^{1/2}} \right )^{1/2} \label{eq2:magnusson1986},
\end{align}
where $(a',b')$ are the projected semiaxes that correspond to the apparent semiaxes of the projected cross section of an ellipsoidal object during a stellar occultation. %Here we are disregarding the orientation $\varphi$ of the projected ellipse.
The rotational light curve amplitude for such an ellipsoid can be calculated as \citep[e.g.,][ p. 426]{Binzel1989}\begin{align}\label{eq1:binzel1989}
\Delta m = 2.5 \log \left(\frac{a}{b}\right) - 1.25 \log \left(\frac{a^2\cos^2(\psi) + c^2\sin^2(\psi)}{b^2\cos^2(\psi) + c^2\sin^2(\psi)}\right).
\end{align}

By performing a grid search for the three body semiaxes, $a,b$, and $c$, and the polar aspect angle, $\psi$, using the rotational phase angle, $\phi,$ for the observed occultation time, we can find the best fit to the projected shape derived from the occultation while simultaneously fitting the rotational light curve amplitude ($\Delta m$). The rotational phase at the time of the observed  2020 October 23 occultation was $\sim~ 0.63$, measured from the absolute maximum of brightness (Fig.~\ref{fig:PhasedRLC}). The observed peak-to-valley amplitude is $\Delta m = 0.36 \pm 0.13$\,mag. We defined the cost function to be minimized as $\chi^2 = (0.36 - \Delta m_c)^ 2 /0.13^2 + (1.5 - a'/b')^ 2 /0.35^2 + (141 - a')^2/9^2 $, with the modeled light-curve amplitude, $\Delta m_c$,  derived by Eq.~\ref{eq1:binzel1989}, and the apparent semiaxes $a',b'$ as obtained from Eqs.~\ref{eq1:magnusson1986}-\ref{eq2:magnusson1986} for each triaxial ellipsoid "clone" (defined by $a,b,c$, and $\psi$; $\phi=0.63\cdot 2\pi$) created during the grid search.
The scanned parameter space was $c=[60,120]$\,km, $b=[c,160]$\,km, and $a=[b,200]$\,km, with a grid spacing of 2\,km. The aspect angle, $\psi,$ was scanned between 0 and 90 degrees in $1\degree$ steps. From this search, we obtained a family of possible solutions of triaxial ellipsoids and aspect angles. The model that minimizes $\chi^2$ has axes $a = 166 \pm 12$\,km, $b = 108 \pm 12$\,km, and $c = 90\substack{+14\\-12}$\,km, with an aspect angle of $\psi = 70\substack{+20\\-12}$~deg. The diameter (of an equal-volume sphere) for this solution is $D_\textrm{eq} = 235 \pm 25$\,km. The $1\sigma$ uncertainties in the retrieved parameters were obtained by varying one parameter from its nominal solution value with corresponding $\chi^2=\chi^2_\textrm{min}$ up to $\chi^2=\chi^2_\textrm{min} + 1$, while keeping the other parameters constant.

The spin axis orientation for this object is currently unknown. But if the pole orientation $(\alpha_p,\delta_p)$ is known, the aspect angle can be computed as
\begin{align}\label{eq:psi_from_pole}
\cos \psi = \sin \delta \sin \delta_p + \cos \delta \cos \delta_0 \cos (\alpha - \alpha_p),
\end{align}
where $\alpha,\delta$ are the coordinates of the object for the date of interest (e.g., the occultation date). By means of a grid search over the whole parameter space in  $1\degree$ steps, we obtained a pole orientation of $(\alpha_p,\delta_p) = (337\degree \pm 10\degree, 62\degree \pm 5 \degree)$.

\subsection{Absolute magnitude and albedo}
From our photometric observations, we obtained an absolute magnitude of $H_R = 5.45 \pm 0.01$\,mag and a phase slope coefficient of $\beta = 0.22 \pm 0.01$\,mag/$\degree$ for \UY{} (Fig.~\ref{fig:R11vsPHA}). Taking into account the brightness contribution due to the rotational phase at the occultation time (about 0.06\,mag), and using a color value $V-R = 0.46 \pm 0.07$\,mag, which we got from our observations as well, this yields  $H_R = 5.51 \pm 0.01$\,mag and $H_V = 5.97 \pm 0.07$\,mag.
Our $V-R$ value is slightly smaller than the $V-R$ value of $0.56 \pm 0.01$\,mag reported by \cite{Sheppard2012} and  the $V-R = 0.59 \pm 0.01$\,mag reported by \cite{Tegler2016}. Sheppard and Tegler derived absolute magnitudes of $H_R = 5.35 \pm 0.03$\,mag and $H_V = 5.91$\,mag, respectively. Taking into account the large light curve amplitude (which was not considered in the estimates derived by these authors), their values are compatible with ours within the error bars. Our result also agrees with the values of $H_R = 5.60 \pm 0.04$\,mag, $H_V = 6.13 \pm 0.04$\,mag, and $V-R = 0.53 \pm 0.06$\,mag obtained by \cite{Alvarez-Candal2019}, where a half rotational light curve amplitude value of $\Delta m / 2 = 0.06$\,mag was also considered in the parameter estimation.
By applying Eq.~(\ref{eq:albmag2diam}) with the values $D_0 = 1330.2$\,km, $D_\mathrm{eq,A} = 228 \pm 21$\,km as area-equivalent diameter, and $H_V = 5.97 \pm 0.07$\,mag, we get a geometric albedo of $p_V = 0.139 \pm 0.027$ for \UY{}.

\subsection{Astrometry}\label{sec:astrometry}
The ellipse center coordinates given in Table~\ref{tab:profile_fit_result}, $(119 \pm 7, -10 \pm 8)$\,km, are two of the five solve-for parameters of the ellipse fit and represent the offset (O-C) between the observed and the predicted position (defined by the object ephemeris and the star position). This information was used to calculate the object position. We derived an astrometric place (ICRS\footnote{International Celestial Reference System}) for \UY{} at 22:19:21.8~UT on 2020 October 23 with equatorial coordinates 
\begin{quote}
\centering
$\alpha$ (hms) = 03 23 26.21404 $\pm$ 0.3 mas,\\
$\delta$ (dms) = +22 47 19.3184 $\pm$ 0.4 mas.
\end{quote}
%
\iffalse
Astrometric object position at time 2020-10-23 22:19:21.800 for reference 'geocenter' (JD 2459146.4301134)
RA = 3 23 26.2140361 +/- 0.324 mas; DEC = 22 47 19.318440 +/- 0.382 mas
\fi
This high-precision astrometry will be used in our orbit determination of the TNO and will also improve the accuracy of future occultation predictions.

\section{Conclusions}\label{sec:conclusions}
\begin{itemize}
\item A stellar occultation by TNO (143707)~\UY{} has been predicted and successfully observed for the first time. From four occultation chords observed at three different sites, we derived an instantaneous projected elliptical size of the object with dimensions $(282 \pm 18) \times (184 \pm 32)$\,km. The area-equivalent diameter is $D_\mathrm{eq,A} = 228 \pm 21$\,km.

\item We also obtained the absolute magnitude, phase slope, $V-R$ color, and rotation period for this TNO from our photometric observation campaign. Our preferred rotation period, $P$, is $12.376 \pm 0.0033$~hours. The light curve is double-peaked with a peak-to-valley amplitude of $\Delta m = 0.36 \pm 0.13$\,mag. We obtained an absolute magnitude of $H_R = 5.51 \pm 0.01$\,mag and $H_V = 5.97 \pm 0.07$\,mag using a $V-R = 0.46 \pm 0.07$\,mag value, which was also derived from our observations. From the area-equivalent diameter of $D_\mathrm{eq,A} = 228 \pm 21$\,km and the  absolute magnitude, $H_V$, given above, we derived a geometric albedo of $p_V = 0.139 \pm 0.027$.

\item By combining the occultation with the rotation light curve results, we derived tight constraints on the three-dimensional size and shape of \UY{}. Our best solution is $2a = 332 \pm 24 $\,km, $2b = 216 \pm 24$\,km, and $2c = 180\substack{+28\\-24}$\,km for a triaxial body, which yields an equivalent spherical diameter of $D_\textrm{eq} = 235 \pm 25$\,km. This value is slightly larger than the radiometric result $D_\textrm{eq} = 196\substack{+114\\-54}$\,km \citep{Farkas-Takacs2020}, but well within the error margins. The aspect angle we derived for the occultation epoch is $\psi = 70\substack{+20\\-12}$\,deg. A pole solution that is compatible with the findings above is $(\alpha_p,\delta_p) = (337\degree \pm 10\degree, 62\degree \pm 5 \degree)$.

\item We derived an occultation-based astrometric position (ICRS) for \UY{}.
\end{itemize}

%%%%%%%%%%%%%%%%%%%%%%%%%%%%%%%%%%%%%%%%%%%%%%%%%%

\begin{acknowledgements}
Part of this work was supported by the Spanish projects PID2020-112789GB-I00 from AEI and Proyecto de Excelencia de la Junta de Andalucía PY20-01309.
Authors JLO, PS-S, NM, MV-L, and RD acknowledge financial support from the Severo Ochoa grant CEX2021-001131-S funded by MCIN/AEI/ 10.13039/501100011033. PS-S also acknowledges financial support from the Spanish I+D+i project PID2022-139555NB-I00 (TNO-JWST) funded by MCIN/AEI/10.13039/501100011033. AA-C acknowledges financial support from the Severo Ochoa grant CEX2021-001131-S funded by MCIN/AEI/10.13039/501100011033. JIBC acknowledges grants 305917/2019-6, 306691/2022-1 (CNPq) and 201.681/2019 (Rio de Janeiro State Research Support Foundation, FAPERJ).
This work is partly based on observations collected at the Centro Astronómico Hispano en Andalucía (CAHA) at Calar Alto, operated jointly by Junta de Andalucía and Consejo Superior de Investigaciones Científicas (CSIC). This research is also partially based on observations carried out at the Observatorio de Sierra Nevada (OSN) operated by Instituto de Astrofísica de Andalucía (IAA-CSIC) and on observations made with the Liverpool Telescope (LT) at the Roque de los Muchachos Observatory (ORM) on the island of La Palma as part of the Observatorios de Canarias (OCAN) operated by the Instituto de Astrofísica de Canarias (IAC).
This study was financed in part by the Coordenação de Aperfeiçoamento de Pessoal de Nível Superior - Brasil (CAPES) - Finance Code 001 and National Institute of Science and Technology of the e-Universe project (INCT do e-Universo, CNPq grant 465376/2014-2). FBR acknowledges CNPq grant 316604/2023-2.
İST60 and İST40 are the observational facilities of the Istanbul University Observatory which were funded by the Scientific Research Projects Coordination Unit of Istanbul University with project numbers BAP-3685 and FBG-2017-23943 and the Presidency of Strategy and Budget of Republic of Turkey with the project 2016K12137.
We acknowledge all observers which might have tried to observe this occultation event and which are not mentioned explicitly in Table~\ref{tab:obs2020}.
This work has made use of data from the European Space Agency (ESA) mission \Gaia\footnote{\scriptsize \url{https://www.cosmos.esa.int/gaia}}, processed by the \Gaia Data Processing and Analysis Consortium (DPAC)\footnote{\scriptsize \url{https://www.cosmos.esa.int/web/gaia/dpac/consortium}}. Funding for the DPAC has been provided by national institutions, in particular the institutions participating in the \Gaia Multilateral Agreement.
We thank the referee and editor  for their valuable comments, which helped us to improve the quality and readability of the manuscript.
This research has made use of the CORA web portal\footnote{\scriptsize \url{https://smallbodies.org/cora}}. 

\end{acknowledgements}

%%%%%%%%%%%%%%%%%%%% REFERENCES %%%%%%%%%%%%%%%%%%

\bibliographystyle{aa}
\bibliography{references}

\providecommand{\noopsort}[1]{}
\begin{thebibliography}{36}
\expandafter\ifx\csname natexlab\endcsname\relax\def\natexlab#1{#1}\fi

\bibitem[{{Alvarez-Candal} {et~al.}(2019){Alvarez-Candal}, {Ayala-Loera},
  {Gil-Hutton}, Ortiz, {Santos-Sanz}, \& Duffard}]{Alvarez-Candal2019}
{Alvarez-Candal}, A., {Ayala-Loera}, C., {Gil-Hutton}, R., {et~al.} 2019,
  Monthly Notices of the Royal Astronomical Society, 488, 3035

\bibitem[{Binzel {et~al.}(1989)Binzel, Gehrels, \& Matthews}]{Binzel1989}
Binzel, R.~P., Gehrels, T., \& Matthews, M.~S. 1989, Asteroids {{II}}

\bibitem[{{Braga-Ribas} {et~al.}(2014){Braga-Ribas}, Sicardy, Ortiz, Snodgrass,
  Roques, {Vieira-Martins}, Camargo, Assafin, Duffard, Jehin, Pollock, Leiva,
  Emilio, Machado, Colazo, Lellouch, Skottfelt, Gillon, Ligier, Maquet,
  {Benedetti-Rossi}, Gomes, Kervella, Monteiro, Sfair, El~Moutamid, Tancredi,
  Spagnotto, Maury, Morales, {Gil-Hutton}, Roland, Ceretta, Gu, Wang,
  Harps{\o}e, Rabus, Manfroid, Opitom, Vanzi, Mehret, Lorenzini, Schneiter,
  Melia, Lecacheux, Colas, Vachier, Widemann, Almenares, Sandness, Char, Perez,
  Lemos, Martinez, J{\o}rgensen, Dominik, Roig, Reichart, Lacluyze, Haislip,
  Ivarsen, Moore, Frank, \& Lambas}]{Braga-Ribas2014}
{Braga-Ribas}, F., Sicardy, B., Ortiz, J.~L., {et~al.} 2014, Nature, 508, 72

\bibitem[{Chandrasekhar(1969)}]{Chandrasekhar1969}
Chandrasekhar, S. 1969, Ellipsoidal Figures of Equilibrium

\bibitem[{Desmars {et~al.}(2015)Desmars, Camargo, {Braga-Ribas},
  {Vieira-Martins}, Assafin, Vachier, Colas, Ortiz, Duffard, Morales, Sicardy,
  {Gomes-J{\'u}nior}, \& {Benedetti-Rossi}}]{Desmars2015}
Desmars, J., Camargo, J. I.~B., {Braga-Ribas}, F., {et~al.} 2015, Astronomy and
  Astrophysics, 584, A96

\bibitem[{{Farkas-Tak{\'a}cs} {et~al.}(2020){Farkas-Tak{\'a}cs}, Kiss,
  Vilenius, Marton, M{\"u}ller, Mommert, Stansberry, Lellouch, Lacerda, \&
  P{\'a}l}]{Farkas-Takacs2020}
{Farkas-Tak{\'a}cs}, A., Kiss, {\relax Cs}., Vilenius, E., {et~al.} 2020,
  Astronomy and Astrophysics, 638, A23

\bibitem[{Ferreira {et~al.}(2022)Ferreira, Tanga, Spoto, Machado, \&
  Herald}]{Ferreira2022}
Ferreira, J.~F., Tanga, P., Spoto, F., Machado, P., \& Herald, D. 2022,
  Astronomy and Astrophysics, 658, A73

\bibitem[{Gault {et~al.}(2022)Gault, Nosworthy, Nolthenius, Bender, \&
  Herald}]{Gault2022}
Gault, D., Nosworthy, P., Nolthenius, R., Bender, K., \& Herald, D. 2022, Minor
  Planet Bulletin, 49, 3

\bibitem[{{Gomes-J{\'u}nior} {et~al.}(2022){Gomes-J{\'u}nior}, Morgado,
  {Benedetti-Rossi}, Boufleur, Rommel, {Banda-Huarca}, Kilic, {Braga-Ribas}, \&
  Sicardy}]{Gomes-Junior2022}
{Gomes-J{\'u}nior}, A.~R., Morgado, B.~E., {Benedetti-Rossi}, G., {et~al.}
  2022, Monthly Notices of the Royal Astronomical Society, 511, 1167

\bibitem[{Harris(1998)}]{Harris1998}
Harris, A.~W. 1998, Icarus, 131, 291

\bibitem[{Horner {et~al.}(2004)Horner, Evans, \& Bailey}]{Horner2004}
Horner, J., Evans, N.~W., \& Bailey, M.~E. 2004, Monthly Notices of the Royal
  Astronomical Society, 354, 798

\bibitem[{Hubbard {et~al.}(1988)Hubbard, Hunten, Dieters, Hill, \&
  Watson}]{Hubbard1988}
Hubbard, W.~B., Hunten, D.~M., Dieters, S.~W., Hill, K.~M., \& Watson, R.~D.
  1988, Nature, 336, 452

\bibitem[{Kaminski {et~al.}(2023)Kaminski, Weber, Marciniak, Zolnowski, \&
  Gedek}]{Kaminski2023}
Kaminski, K., Weber, C., Marciniak, A., Zolnowski, M., \& Gedek, M. 2023, PASP,
  135, 025001

\bibitem[{Kervella {et~al.}(2004)Kervella, Th{\'e}venin, Di~Folco, \&
  S{\'e}gransan}]{Kervella2004}
Kervella, P., Th{\'e}venin, F., Di~Folco, E., \& S{\'e}gransan, D. 2004,
  Astronomy and Astrophysics, 426, 297

\bibitem[{Kilic {et~al.}(2022)Kilic, {Braga-Ribas}, Kaplan, Erece, Souami,
  Dindar, Desmars, Sicardy, Morgado, Shameoni, Rommel, \&
  {Gomes-J{\'u}nior}}]{Kilic2022}
Kilic, Y., {Braga-Ribas}, F., Kaplan, M., {et~al.} 2022, Monthly Notices of the
  Royal Astronomical Society, 515, 1346

\bibitem[{Leiva {et~al.}(2020)Leiva, Buie, Keller, Wasserman, Kavelaars,
  Bridges, Haley, Strauss, Wilde, Weryk, Kervella, Baker, Bock, Conway, Cota,
  Estes, Garc{\'i}a, Kehrli, McCandless, McCandless, Self, Settlemire, Swanson,
  Thompson, \& Wise}]{Leiva2020}
Leiva, R., Buie, M.~W., Keller, J.~M., {et~al.} 2020, Planet. Sci. J., 1, 48

\bibitem[{Lellouch {et~al.}(2013)Lellouch, {Santos-Sanz}, Lacerda, Mommert,
  Duffard, Ortiz, M{\"u}ller, Fornasier, Stansberry, Kiss, Vilenius, Mueller,
  Peixinho, Moreno, Groussin, Delsanti, \& Harris}]{Lellouch2013}
Lellouch, E., {Santos-Sanz}, P., Lacerda, P., {et~al.} 2013, Astronomy and
  Astrophysics, 557, A60

\bibitem[{Lomb(1976)}]{Lomb1976}
Lomb, N.~R. 1976, Astrophysics and Space Science, 39, 447

\bibitem[{Magnusson(1986)}]{Magnusson1986}
Magnusson, P. 1986, Icarus, 68, 1

\bibitem[{Morales {et~al.}(2022)Morales, Ortiz, Morales,
  {Fern{\'a}ndez-Valenzuela}, {Santos-Sanz}, Duffard, Kretlow, \&
  Vara}]{Morales2022}
Morales, N., Ortiz, J.~L., Morales, R., {et~al.} 2022, EPSC2022

\bibitem[{M{\"u}ller {et~al.}(2009)M{\"u}ller, Lellouch, B{\"o}hnhardt,
  Stansberry, Barucci, Crovisier, Delsanti, Doressoundiram, Dotto, Duffard,
  Fornasier, Groussin, Guti{\'e}rrez, Hainaut, Harris, Hartogh, Hestroffer,
  Horner, Jewitt, Kidger, Kiss, Lacerda, Lara, Lim, Mueller, Moreno, Ortiz,
  Rengel, {Santos-Sanz}, Swinyard, Thomas, Thirouin, \& Trilling}]{Muller2009}
M{\"u}ller, T.~G., Lellouch, E., B{\"o}hnhardt, H., {et~al.} 2009, Earth Moon
  Planet, 105, 209

\bibitem[{Nesvorn{\'y} \& Morbidelli(2012)}]{Nesvorny2012}
Nesvorn{\'y}, D. \& Morbidelli, A. 2012, The Astronomical Journal, 144, 117

\bibitem[{Oliveira {et~al.}(2022)Oliveira, Sicardy, {Gomes-J{\'u}nior}, Ortiz,
  Strobel, Bertrand, Forget, Lellouch, Desmars, B{\'e}rard, Doressoundiram,
  Lecacheux, Leiva, Meza, Roques, Souami, Widemann, {Santos-Sanz}, Morales,
  Duffard, {Fern{\'a}ndez-Valenzuela}, {Castro-Tirado}, {Braga-Ribas}, Morgado,
  Assafin, Camargo, {Vieira-Martins}, {Benedetti-Rossi}, {Santos-Filho},
  {Banda-Huarca}, {Quispe-Huaynasi}, Pereira, Rommel, Margoti, {Dias-Oliveira},
  Colas, Berthier, Renner, Hueso, {P{\'e}rez-Hoyos}, {S{\'a}nchez-Lavega},
  Rojas, Beisker, Kretlow, Herald, Gault, Bath, Bode, Bredner, Guhl, Haymes,
  Hummel, Kattentidt, Kl{\"o}s, Pratt, Thome, Avdellidou, Gazeas, Karampotsiou,
  Tzouganatos, Kardasis, Christou, Xilouris, Alikakos, Gourzelas, Liakos,
  Charmandaris, Jel{\'i}nek, {\v S}trobl, Eberle, Rapp, G{\"a}hrken, Klemt,
  Kowollik, Bitzer, Miller, Herzogenrath, Frangenberg, Brandis, P{\"u}tz,
  Perdelwitz, Piehler, Riepe, {\noopsort{poschinger}}von Poschinger,
  Baruffetti, Cenadelli, Christille, Ciabattari, Luca, Alboresi, Leto, Sanchez,
  Bruno, Occhipinti, Morrone, Cupolino, Noschese, Vecchione, Scalia, Savio,
  Giardina, Kamoun, Barbosa, Behrend, Spano, Bouchet, Cottier, Falco, Gallego,
  Tortorelli, Sposetti, Sussenbach, Abbeel, Andr{\'e}, Llibre, Pailler,
  Ardissone, Boutet, Sanchez, Bretton, Cailleau, Pic, Granier, Chauvet, Conjat,
  Dauvergne, Dechambre, Delay, Delcroix, Rousselot, Ferreira, Machado, Tanga,
  Rivet, Frappa, Irzyk, Jabet, Kaschinski, Klotz, Rieugnie, Klotz, Labrevoir,
  Lavandier, Walliang, Leroy, Bouley, Lisciandra, Coliac, Metz, Erpelding,
  Nougayr{\`e}de, Midavaine, Miniou, Moindrot, Morel, Reginato, Reginato,
  Rudelle, Tregon, Tanguy, David, Thuillot, Hestroffer, Vaudescal, Aissa,
  Grigahcene, Briggs, Broadbent, Denyer, Haigh, Quinn, Thurston, Fossey, Arena,
  Jennings, Talbot, Alonso, Reche, Casanova, Briggs, {Iglesias-Marzoa},
  Ib{\'a}{\~n}ez, Mart{\'i}n, Gonz{\'a}lez, Garc{\'i}a, Marchant,
  {Ordonez-Etxeberria}, Martorell, Salamero, Organero, Ana, Fonseca, Peris,
  Brevia, Selva, Perello, Cabedo, Gon{\c c}alves, Ferreira, Dias, Daassou,
  Barkaoui, Benkhaldoun, Guennoun, Chouqar, Jehin, Rinner, Lloyd, Moutamid,
  Lamarche, Pollock, Caton, Kouprianov, Timerson, Blanchard, Payet, Peyrot,
  {Teng-Chuen-Yu}, Fran{\c c}oise, Mondon, Payet, Boissel, Castets, Hubbard,
  Hill, Reitsema, Mousis, Ball, Neilsen, Hutcheon, Lay, Anderson, Moy, Jonsen,
  Pink, Walters, \& Downs}]{Oliveira2022}
Oliveira, J.~M., Sicardy, B., {Gomes-J{\'u}nior}, A.~R., {et~al.} 2022, A\&A,
  659, A136

\bibitem[{Ortiz {et~al.}(2015)Ortiz, Duffard, {Pinilla-Alonso},
  {Alvarez-Candal}, {Santos-Sanz}, Morales, {Fern{\'a}ndez-Valenzuela},
  Licandro, Campo~Bagatin, \& Thirouin}]{Ortiz2015}
Ortiz, J.~L., Duffard, R., {Pinilla-Alonso}, N., {et~al.} 2015, Astronomy and
  Astrophysics, 576, A18

\bibitem[{Ortiz {et~al.}(2017)Ortiz, {Santos-Sanz}, Sicardy, {Benedetti-Rossi},
  B{\'e}rard, Morales, Duffard, {Braga-Ribas}, Hopp, Ries, Nascimbeni, Marzari,
  Granata, P{\'a}l, Kiss, Pribulla, Kom{\v z}{\'i}k, Hornoch, Pravec, Bacci,
  Maestripieri, Nerli, Mazzei, Bachini, Martinelli, Succi, Ciabattari, Mikuz,
  Carbognani, Gaehrken, Mottola, Hellmich, Rommel, {Fern{\'a}ndez-Valenzuela},
  Bagatin, Cikota, Cikota, Lecacheux, {Vieira-Martins}, Camargo, Assafin,
  Colas, Behrend, Desmars, Meza, {Alvarez-Candal}, Beisker, {Gomes-Junior},
  Morgado, Roques, Vachier, Berthier, Mueller, Madiedo, Unsalan, Sonbas,
  Karaman, Erece, Koseoglu, Ozisik, Kalkan, Guney, Niaei, Satir, Yesilyaprak,
  Puskullu, Kabas, Demircan, Alikakos, Charmandaris, Leto, Ohlert, Christille,
  Szak{\'a}ts, Farkas, {Varga-Vereb{\'e}lyi}, Marton, Marciniak, Bartczak,
  {Santana-Ros}, {Butkiewicz-B{\k{a}}k}, Dudzi{\'n}ski, {Al{\'i}-Lagoa},
  Gazeas, Tzouganatos, Paschalis, Tsamis, {S{\'a}nchez-Lavega},
  {P{\'e}rez-Hoyos}, Hueso, Guirado, Peris, \& {Iglesias-Marzoa}}]{Ortiz2017}
Ortiz, J.~L., {Santos-Sanz}, P., Sicardy, B., {et~al.} 2017, Nature, 550, 219

\bibitem[{Rieke {et~al.}(2008)Rieke, Blaylock, Decin, Engelbracht, Ogle,
  Avrett, Carpenter, Cutri, Armus, Gordon, Gray, Hinz, Su, \&
  Willmer}]{Rieke2008}
Rieke, G.~H., Blaylock, M., Decin, L., {et~al.} 2008, AJ, 135, 2245

\bibitem[{Rommel {et~al.}(2020)Rommel, {Braga-Ribas}, Desmars, Camargo, Ortiz,
  Sicardy, {Vieira-Martins}, Assafin, {Santos-Sanz}, Duffard,
  {Fern{\'a}ndez-Valenzuela}, Lecacheux, Morgado, {Benedetti-Rossi},
  {Gomes-J{\'u}nior}, Pereira, Herald, Hanna, Bradshaw, Morales, Brimacombe,
  Burtovoi, Carruthers, {\noopsort{barros}}{de Barros}, Fiori, Gilmore, Hooper,
  Hornoch, Jacques, Janik, Kerr, Kilmartin, Winkel, Naletto, Nardiello,
  Nascimbeni, Newman, Ossola, P{\'a}l, Pimentel, Pravec, Sposetti, Stechina,
  Szak{\'a}ts, Ueno, Zampieri, Broughton, Dunham, Dunham, Gault, Hayamizu,
  Hosoi, Jehin, Jones, Kitazaki, Kom{\v z}{\'i}k, Marciniak, Maury, Miku{\v z},
  Nosworthy, F{\'a}brega~Polleri, Rahvar, Sfair, Siqueira, Snodgrass, Sogorb,
  Tomioka, {Tregloan-Reed}, \& Winter}]{Rommel2020}
Rommel, F.~L., {Braga-Ribas}, F., Desmars, J., {et~al.} 2020, Astronomy and
  Astrophysics, 644, A40

\bibitem[{Russell(1916)}]{Russell1916}
Russell, H.~N. 1916, The Astrophysical Journal, 43, 173

\bibitem[{{Santos-Sanz} {et~al.}(2012){Santos-Sanz}, Lellouch, Fornasier, Kiss,
  Pal, M{\"u}ller, Vilenius, Stansberry, Mommert, Delsanti, Mueller, Peixinho,
  Henry, Ortiz, Thirouin, Protopapa, Duffard, Szalai, Lim, Ejeta, Hartogh,
  Harris, \& Rengel}]{Santos-Sanz2012}
{Santos-Sanz}, P., Lellouch, E., Fornasier, S., {et~al.} 2012, Astronomy and
  Astrophysics, 541, A92

\bibitem[{Sarid {et~al.}(2019)Sarid, Volk, Steckloff, Harris, Womack, \&
  Woodney}]{Sarid2019}
Sarid, G., Volk, K., Steckloff, J.~K., {et~al.} 2019, The Astrophysical Journal
  Letters, 883, L25

\bibitem[{Scargle(1982)}]{Scargle1982}
Scargle, J.~D. 1982, The Astrophysical Journal, 263, 835

\bibitem[{Sheppard(2012)}]{Sheppard2012}
Sheppard, S.~S. 2012, The Astronomical Journal, 144, 169

\bibitem[{Sicardy {et~al.}(2003)Sicardy, Widemann, Lellouch, Veillet,
  Cuillandre, Colas, Roques, Beisker, Kretlow, Lagrange, Gendron, Lacombe,
  Lecacheux, Birnbaum, Fienga, Leyrat, Maury, Raynaud, Renner, Schultheis,
  Brooks, Delsanti, Hainaut, Gilmozzi, Lidman, Spyromilio, Rapaport,
  Rosenzweig, Naranjo, Porras, D{\'i}az, Calder{\'o}n, Carrillo, Carvajal,
  Recalde, Cavero, Montalvo, Barr{\'i}a, Campos, Duffard, \&
  Levato}]{Sicardy2003}
Sicardy, B., Widemann, T., Lellouch, E., {et~al.} 2003, Nature, 424, 168

\bibitem[{Tegler {et~al.}(2016)Tegler, Romanishin, Consolmagno, \&
  J.}]{Tegler2016}
Tegler, S.~C., Romanishin, W., Consolmagno, G.~J., \& J., S. 2016, The
  Astronomical Journal, 152, 210

\bibitem[{Willmer(2018)}]{Willmer2018}
Willmer, C. N.~A. 2018, ApJS, 236, 47

\bibitem[{Zacharias {et~al.}(2004)Zacharias, Monet, Levine, Urban, Gaume, \&
  Wycoff}]{Zacharias2004}
Zacharias, N., Monet, D.~G., Levine, S.~E., {et~al.} 2004, 205, 48.15

\end{thebibliography}

%%%%%%%%%%%%%%%%% APPENDICES %%%%%%%%%%%%%%%%%%%%%

\begin{appendix}
%\section{}
\setcounter{section}{1}
\begin{table*}
\caption{Observation details for the 2020 October 23 occultation.}\label{tab:obs2020}
\vspace{-0.25em}
\centering
\footnotesize
\begin{tabular}{@{\extracolsep{\fill}}rllllllllll@{}}
%\hline\hline
\toprule\midrule
\# & Site name          & CC &  Latitude (dms)          & Telescope (cm)         & Method & Observation \\ & Observer(s)         &    &  Longitude (dms) & Camera & ExpTime &   DeadTime \\  &   &   &  Elevation (m)  &  &  TimeSrc &   \\\midrule
1  & El Arenosillo                                      & ES & N 37 05 53.0& T35     & IMG   & Positive  \\& & & W 06 44 06.0 & Andor iXon DV-897-BV & 7.0\,s& 0.567\,s\\ & \multicolumn{1}{l}{\parbox[t]{5.5cm}{\raggedright\em E. J. Fernández-García, A. Castro-Tirado}} & &  40 &  &  NTP &\\\midrule
2  & Granada                                                    & ES & N 37 06 41.0& T25         & IMG   & Positive  \\& & & W 03 38 22.1 & ZWO ASI178M & 2.0\,s        & 0.061\,s\\    & \multicolumn{1}{l}{\parbox[t]{5cm}{\raggedright\em M. Sanchez}} & &  730 &  &  NTP &\\\midrule
3  & Sierra Nevada Observatory (OSN90)  & ES & N 37 03 46.40& T90       & IMG     & Positive  \\& & & W 03 23 09.35 & Roper Scientific VersArray 2048B & 3.0\,s & 2.181\,s\\ & \multicolumn{1}{l}{\parbox[t]{5cm}{\raggedright\em A. Sota, PI: P. Santos-Sanz, J. Ortiz}} & &  2841 &  &  NTP &\\\midrule
4  & Sierra Nevada Observatory (OSN150) & ES & N 37 03 46.52& T150      & IMG     & Positive  \\& & & W 03 23 09.67       & Andor iKon-L 936 BEX2-DD & 3.0\,s & 1.848\,s \\ & \multicolumn{1}{l}{\parbox[t]{5cm}{\raggedright\em A. Sota, PI: P. Santos-Sanz, J. Ortiz}} & &  2841 &  &  NTP &\\\midrule%\midrule
5  & Leeds                                              & UK & N 53 50 15.4 & T28   & VID   & Negative  \\& & & W 01 36 28.0 &  Watec 910HX & 0.64\,s & - \\ & \multicolumn{1}{l}{\parbox[t]{5cm}{\raggedright\em A. Pratt}} & &  114 &  & GPS &\\\midrule
6  & Stevenage                                  & UK & N 51 57 04.4 & T35         & IMG   & Negative  \\& & & W 00 03 51.7 &  ZWO ASI174 mono & 0.34\,s & - \\ & \multicolumn{1}{l}{\parbox[t]{5cm}{\raggedright\em S. Kidd}} & &  120 &  & GPS &\\\midrule
7  & Oxford                                         & UK & N 51 55 41.2 & T28     & IMG   & Negative  \\& & & W 01 18 46.3 &  QHY174m-GPS & 1.0\,s & - \\ & \multicolumn{1}{l}{\parbox[t]{5cm}{\raggedright\em T. Haymes}} & &  199 &  & GPS &\\\midrule
8  & Abingdon                                   & UK & N 51 37 53.1 & T30         & VID   & Negative  \\& & & W 01 16 55.2 &  Watec 910HX & 0.64\,s & - \\ & \multicolumn{1}{l}{\parbox[t]{5cm}{\raggedright\em J. C. Talbot}} & &  59 &  & GPS &\\\midrule
9  & Dourbes                                        & BE & N 50 05 25.9 & T40   & VID     & Negative  \\& & & E 04 34 56.0 & Watec 910 HX/RC & 0.32\,s & - \\ & \multicolumn{1}{l}{\parbox[t]{5cm}{\raggedright\em R. Boninsegna}} & &  195 &  &  GPS &\\\midrule
10  & Cannet                                        & FR & N 43 37 15.2 & T40   & VID     & Negative  \\& & & W 00 02 40.7 & Watec 120N+ & 1.28\,s & - \\ & \multicolumn{1}{l}{\parbox[t]{5cm}{\raggedright\em J. J. Castellani}} & &  180 &  &  GPS &\\\midrule
11 & Istanbul University Observatory    & TR & N 41 00 42.48& T40   & IMG   & Negative  \\& Application and Research Center (İST40) & & E 28 57 56.34 & Moravian G2& 0.9\,s  & - \\ & \multicolumn{1}{l}{\parbox[t]{5cm}{\raggedright\em S. Fişek, O. Çakır}} & &  60 &  &  NTP &\\\midrule
12  & Ibiza                                             & ES & N 38 53 28.0 & T50   & IMG   & Negative  \\& & & E 01 14 26.9 & SBIG STL-11000 & 3\,s & 3\,s \\ & \multicolumn{1}{l}{\parbox[t]{5cm}{\raggedright\em I. de la Cueva}} & &  166 &  &  NTP &\\\midrule
13  & Albox                                             & ES & N 37 24 20.0& T40     & IMG   & Negative  \\& & & W 02 09 06.5 & Atik 314L+ & 2.5\,s & 0.8 \\ & \multicolumn{1}{l}{\parbox[t]{5cm}{\raggedright\em J. L. Maestre}} & &  493 &  &  NTP &\\\midrule
14 & Calar Alto Observatory (CAHA220)   & ES & N 37 13 23.30& T220  & IMG   & Negative  \\ & & & W 02 32 46.30 & CAFOS autoguider camera & 7\,s & 2\,s \\ & \multicolumn{1}{l}{\parbox[t]{5.5cm}{\raggedright\em A. Guijarro, PI: J. Ortiz, P. Santos-Sanz}} & &  2168 &  &  NTP &\\\midrule
15 & Las Negras                                     & ES & N 36 52 50.0 & T20   & IMG     & Negative  \\& & & W 02 00 54.0  & CCD & 10.0\,s & 2.5\,s \\ & \multicolumn{1}{l}{\parbox[t]{5cm}{\raggedright\em F. Casarramona}} & &  50 &  &  NTP &\\\midrule
16 & Teide Observatory (MUSCAT2)            & ES & N 28 28 28.6& T150   & IMG     & Negative  \\& & & W 16 18 29.0  & CCD & 20\,s & 3\,s \\ & \multicolumn{1}{l}{\parbox[t]{5cm}{\raggedright\em E. Pallé, F. Murgas}} & & 2380 &  &  NTP &\\\midrule
17 & Teide Observatory (TAR1 \& TAR2)   & ES & N 28 28 28.6& T46        & IMG     & Negative  \\& & & W 16 18 29.0  & CCD & 0.5\,s & 0.4\,ms \\ & \multicolumn{1}{l}{\parbox[t]{5cm}{\raggedright\em M. R. Alarcon, J. Licandro, M. Serra-Ricart}} & & 2380 &  &  NTP &\\\midrule
\end{tabular}
\tablefoot{\scriptsize
CC is the two-letter country code. Site latitude, longitude (format dms) and elevation (AMSL in m) are given in the WGS84 datum. Telescope: Tx is the telescope aperture in cm. Method is the recording method: IMG means digital (CDD, CMOS) sequential imaging, VID is analog video recording. TimeSrc is the used timing source and method: GPS (Global Positioning System) means 1-PPS (one pulse per second) driven video-time-insertion (VID) or camera-internal GPS timestamps (IMG). NTP denotes a Network Time Protocol computer system clock synchronization. Observation: either positive (occultation detected/recorded) or negative. ExpTime is the exposure time in seconds, DeadTime is the dead time in seconds between two subsequent images during the recording.}
\end{table*}
\end{appendix}
\end{document}